\documentclass[12pt,preprint]{aastex}

\newcommand{\lsim}{\raise0.3ex\hbox{$<$}\kern-0.75em{\lower0.65ex\hbox{$\sim$}}}
\newcommand{\gsim}{\raise0.3ex\hbox{$>$}\kern-0.75em{\lower0.65ex\hbox{$\sim$}}}
\newcommand{\propsim}{\raise0.3ex\hbox{$\propto$}\kern-0.75em{\lower0.65ex\hbox{$\sim$}}}

\begin{document}
    
\title{Inhomogeneities and the modeling of radio supernovae}

\author{C.-I. Bj\"ornsson\altaffilmark{1} and S.T. Keshavarzi\altaffilmark{1}}
\altaffiltext{1}{Department of Astronomy, AlbaNova University Center, Stockholm University, SE--106~91 Stockholm, Sweden.}
\email{bjornsson@astro.su.se}

\begin{abstract}
Observations of radio supernovae often exhibit characteristics not readily accounted for by a homogeneous, spherically symmetric synchrotron model; e.g., flat-topped spectra/lightcurves. It is shown that many of these deviations from the standard model can be attributed to an inhomogeneous source structure. When inhomogeneities are present, the deduced radius of the source and, hence, the shock velocity, is sensitive to the details of the modeling. Since the inhomogeneities are likely to result from the same mechanism that amplify the magnetic field, a comparison between observations and the detailed numerical simulations now under way may prove mutually beneficial. It is argued that the radio emission in Type Ib/c supernovae has a small volume filling factor and comes from a narrow region associated with the forward shock, while the radio emission region in SN 1993J (Type IIb) is determined by the extent of the Rayleigh-Taylor instability emanating from the contact discontinuity. Attention is also drawn to the similarities between radio supernovae and the structural properties of supernova remnants.
\end{abstract}


\section{Introduction}
The radio emission observed from supernovae (SNe) is generally agreed to be synchrotron radiation. For most radio supernovae, the emission is not spatially resolved so the source radius, $R$, has to be deduced from modeling. In the standard model \citep{che82a}, the interaction between supernova ejecta and circumstellar medium gives rise to two shocks. The radio emission is thought to come from the region in between these shocks. The interplay between circumstellar medium and ejecta determines the variation of the radius of the forward shock with time ($t$) as $R \propto t^{(n-3)/(n-2)}$, where $n$ reflects the properties of the ejecta. Since the model assumes spherical symmetry and a homogeneous source, radio observations give estimates of the density of the circumstellar medium and the ejecta structure. However, this assumes that the thermal energy density behind the shocks can be related to the radio derived energy densities of magnetic fields and relativistic electrons. Unfortunately, this is not the case. Hence, for example, the deduced mass loss rate of the progenitor star depends sensitively on the values of these unknown free parameters.

Another prerequisite for using radio data to derive supernovae properties is that of a good model fit. This is indeed often the case for the standard model. However, with the increased quality of radio observations, it has become evident that deviations from the standard model do occur. Various attempts have been made to modify the standard model to account for these effects. Often, such additions have had little physical underpinning. This paper considers the possibility that inhomogeneities in an otherwise spherically symmetric source can provide a physical starting point for an extension of the standard model. 

The presence of inhomogeneities would affect several of the conclusions drawn from observations using the standard model. The most direct one concerns the radius for a spatially unresolved source. Since the velocity of the forward shock is deduced from the value of $R$, this would impact, for example, the discussion of the existence of a central engine \citep{k/b10,sod10}. The strength and connection to other supernova properties of such an extra source of energy could potentially be related to the origin of superluminous supernovae \citep{qui11,gal12}. Furthermore, as discussed in \cite{bjo13}, inhomogeneities can affect the expected emission in other wavelength ranges; for example, the inverse Compton scattered radiation may increase relative the radio emission.

The properties of these inhomogeneities are likely determined by the same processes that amplify the magnetic field and possibly also accelerate the relativistic electrons. Inhomogeneities may then provide a direct link between observations and the processes shaping the non-thermal aspects of the shocked gas. 

It is thought that the magnetic field strength results from a turbulent dynamo in which part of the kinetic turbulent energy is converted into magnetic energy. The efficiency of this mechanism is still uncertain \citep[e.g.,][]{sch04}. Furthermore, there are two qualitatively different settings for this scenario. In the local, small-scale dynamo, the amplified magnetic field is isotropic. Spatially unresolved radio supernovae show little polarization. This could be caused by either an isotropic magnetic field or a spherically symmetric source geometry. In this connection, one may note that SN 1572 (Tycho) is spatially resolved and shows substantially polarized radio emission \citep{dic91}. As is argued later in this paper, the deduced properties of the inhomogeneities support the presence of such a large scale magnetic field structure also in the spatially unresolved sources. 

This leaves a global, large-scale dynamo, where the magnetic field results from a combination of a large scale anisotropy and turbulence, as the agent for magnetic field amplification. The pressure gradient between the two supernova shocks gives rise to turbulence emanating from the contact discontinuity \citep{che92}. Magnetohydrodynamic (MHD) simulations show that this in turn generates a magnetic field with a large scale structure \citep{j/n96a,j/n96b}. Another place where turbulence is expected to occur in an anisotropic background is the region around the shock front. It is interesting to note that detailed numerical simulations of shock physics have become possible in recent years. Such particle-in-cell (PIC) simulations treat the amplification of the magnetic field and the acceleration of particles self-consistently as part of the shock formation process \citep{c/s14a,c/s14b}. Although the scope of these simulations is still very limited, they provide indications of the properties of the inhomogeneities to be expected in future, more extensive calculations. 

The aim of this paper is two-fold: Firstly, to argue that inhomogeneities can account for many of the shortcomings of the standard model made evident by detailed radio observations. Secondly, while simulations may help to convert radio observations into physical supernova properties, the reverse is also likely to be true; namely, detailed radio observations could be used to reveal and constrain the physical mechanisms at work in the supernova shock region. Section \ref{sect2} discusses a few well-known supernovae. They have been chosen in order to illustrate how various methods to accomodate the standard model to observations affect the deduced source parameters. A simple way to introduce inhomogeneities into the standard model is presented in section \ref{sect3}. Instead of attempting a detailed fit to observations, it is shown how a rough estimate of a few crucial parameters describing the inhomogeneities can be obtained. A discussion follows in section \ref{sect4}. The focus is on the two mechanisms discussed above for the amplification of the magnetic field and how observations can be used to distinguish between them. The conclusions of the paper are collected in section \ref{sect5}.

\section{The validity of the standard model} \label{sect2}

In a homogeneous, spherically symmetric synchrotron source, in addition to the optically thin spectral index, the spectrum is determined by the synchrotron self-absorption frequency ($\nu_{\rm abs}$) and the corresponding spectral flux ($F(\nu_{\rm abs})$). The source properties are specified by three physical quantities; namely, the magnetic field strength ($B$), the energy density of relativistic electrons ($U_{\rm e}$), and the radius of the source ($R$). The relations between the source parameters and the observed quantities can be written (see Appendix)
\begin{eqnarray}
	BR & \propto &  F(\nu_{\rm abs})^{\frac{p+4}{2p+13}}\, \hat{y}^{\frac{-5}{2p+13}} \nonumber\\
	B & \propto &  \nu_{\rm abs}\, F(\nu_{\rm abs})^{\frac{-2}{2p+13}}\, \hat{y}^{\frac{-4}{2p+13}},
	\label{eq3}
\end{eqnarray}
where
\begin{equation}
	\hat{y} \propto \frac{U_{\rm e}}{U_{\rm B}}\frac{r}{R}\, \gamma_{\rm min}^{p-2}.
	\label{eq4}
\end{equation}
The electron energy distribution is here given as $N(\gamma) \propto \gamma^{-p}$ for $\gamma > \gamma_{\rm min}$ and $p>2$. Furthermore, $r$ is the radial depth of the emission zone. In order to solve for the source parameters, a value for $\hat{y}$ is needed. It is often assumed that the energy density of the magnetic field $U_{\rm B} (\equiv B^2/8\pi) = U_{\rm e}$, i.e., that equipartition prevails.

There are potentially other observables that could be used to deduce the source properties; for example, in a spatially resolved source, $R$ is readily obtained or if cooling is important, either due to synchrotron or inverse Compton radiation, the value of $B$ can be determined. However, only in rare cases are any of these latter observables available. It is, therefore, of interest to determine to what extent conclusions drawn from the standard model apply to real supernovae or whether they are the result of  inappropriate modeling. Below a few issues of modeling radio supernovae are discussed. The examples have been chosen in order to address two aspects of the modeling; firstly, the applicability of the standard model  and, secondly, when inappropriate, how much does its use affect the derived source properties. 

\subsection{SN 1994I}\label{sect2aa}

The radio emission from SN 1994I was extensively observed by \cite{wei11}. These observations have been discussed by \cite{ale15} using the standard model. In this model, the peak spectral flux is related by a constant factor to the peak flux of the light curve for the corresponding frequency (note, for convenience, the same notation is used below for both of these fluxes). 

In order to fit the rising, optically thick part of the light curves, \cite{ale15} assume a radius varying as $R \propto t^{0.88}$. However, the deduced variation of  $\nu_{\rm abs}$ then results in an accelerated flow, which is inconsistent both with the initial assumption and the standard model. This contradiction cannot be circumvented by choosing another temporal dependence for the source radius, as can be seen from the following argument: The observed light curves imply a rapidly decreasing self-absorption frequency ($\nu_{\rm abs}\, \propsim\, t^{-1.2}$). The rising part of a light curve is obtained from $F_{\nu}(t) \propto R^2 B^{-1/2}$ or $F_{\nu}(t) \propto \nu_{\rm abs}^{-5/2}\, F(\nu_{\rm abs})$ (cf. eqn (\ref{eq3})). Since $F(\nu_{\rm abs})$ is roughly constant, $F_{\nu}(t) \propto t^3$ is expected. Instead, a substantially lower rate of increase is found, $F_{\nu}(t) \propto t^{2.3}$. Such a discrepancy is larger than allowed by the measurement errors. 

 Further indications of shortcomings of the standard model come from the spectra. Although they are not as well sampled as the light curves, the source properties derived differ appreciably with those from the light curves. A reason for these discrepant results, when using the standard model, may be suggested by the peak morphology of the light curves. There is a tendency for the peaks to be somewhat flat-topped, in particularly, for the early, high frequency light curves (i.e., 8.4 GHz and 15.0 GHz). The value of $\nu_{\rm abs}$ and its evolution with time would then be less well defined.

\subsection{SN 2011dh}\label{sectab} 

The radio observations of SN 2011dh were presented in \cite{sod12} and \cite{kra12}. They found $\nu_{\rm abs}\, \propsim\, t^{-0.9}$, which together with a roughly constant $F(\nu_{\rm abs})$, gave a physically plausible fit to the observations. However, in order to get reduced $\chi^2$-values of unity, they needed to artificially increase their measurement errors by a factor of 3\,-\,7. This was required, since their fits showed systematic deviations from the standard model. Firstly, there was excess emission below $\nu_{\rm abs}$ and, secondly, the derived value of $\nu_{\rm abs}$ was lower than the apparent peak in the spectra (i.e., excess emission also above $\nu_{\rm abs}$). Both of these features are reminiscent of those found in SN 1994I, i.e., spectra/light curves more flat-topped than expected in the standard model.

Independent of the origin of the excess emission, it is clear that the radius must be larger than deduced using the standard model. SN 2011dh is one of a few radio supernovae, which has been spatially resolved by VLBI \citep{bie12}. This makes it possible to directly measure the source radius. Using the radii deduced from the standard model during the earlier phases of the expansion, \cite{dew16} find that the radius increases according to $n\approx 26$. However, this large value of $n$ is sensitive to the accuracy of the radii deduced for the initial expansion. The consistent appearance of excess emission during this phase and the resulting underestimation of the source size, suggest that this $n$-value should be regarded as an upper limit. 

\subsection{PTF11qcj}\label{sectac}

\begin{figure}
\plotone{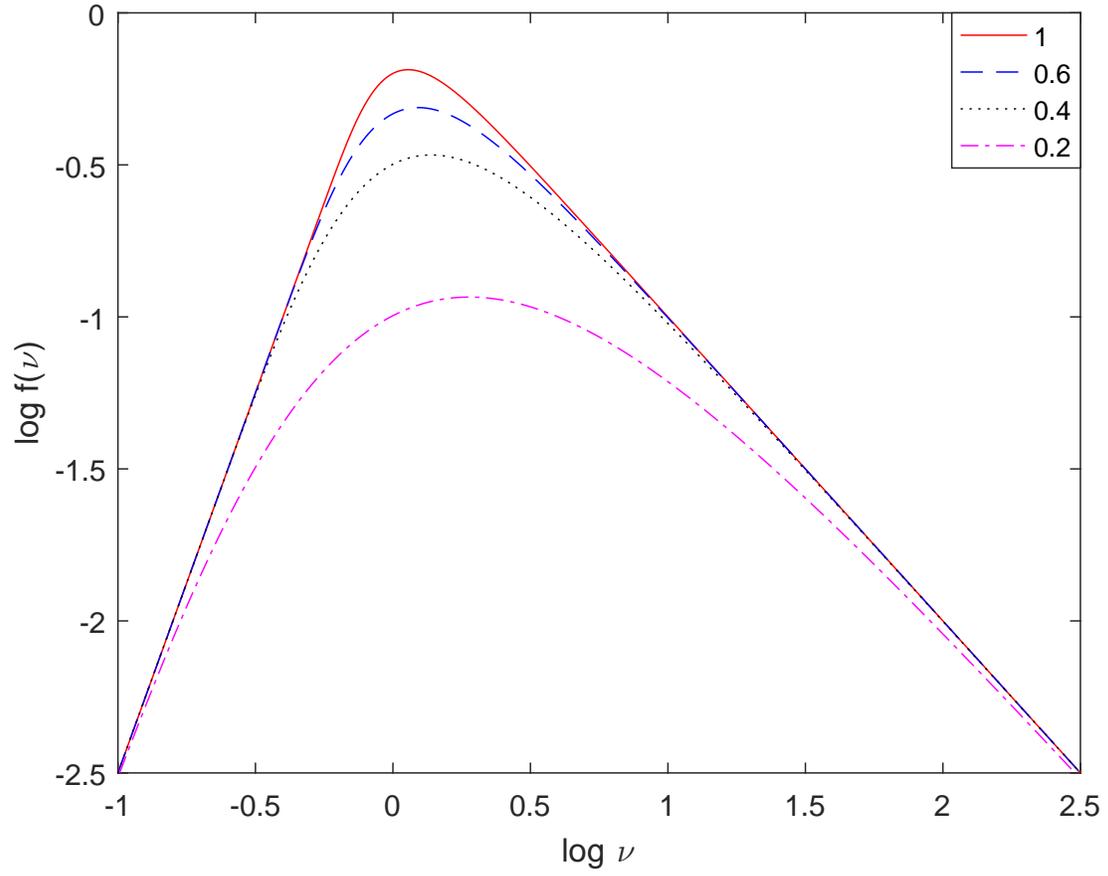}
\caption{The local synchrotron emissivity modified by the parameter $\xi$ (see text). $\xi = 1$ corresponds to the standard emissivity.
\label{fig1}} 
\end{figure}

The presence of flat-topped spectra/light curves has been recognized for some time. Various methods have been introduced in order to use the standard model as a starting point for the analysis. The increase of the measurement errors was discussed above. An alternative way was suggested by \cite{sod05}. The frequency distribution of the synchrotron radiation was artificially broadened by introducing a parameter $\xi$, which "cuts off" the top of the locally emitted spectrum according to (see also figure \ref{fig1})
\begin{equation}
	f_{\xi}(\nu) \propto \frac{\nu^{5/2}}{B^{1/2}} \left[1-\exp \left(-\tau^{\xi}(\nu)\right)\right]^{1/\xi},
	\label{eq5}
\end{equation}
where $\tau(\nu) = (\nu/\nu_{\rm abs})^{-(p+4)/2}$ is the optical depth. The standard synchrotron spectrum corresponds to $\xi = 1$. Various values for $\xi$ have been derived; for example, $\xi \approx 0.6$ for SN\,2003BG \citep{sod06}, $\xi \approx 0.5$ for SN\,2003L \citep{sod05}, and the most extreme one, $\xi \approx 0.2$ for PTF11qcj \citep{cor14}.

Normally, no renormalization of the local emissivity is done. Hence, the derived radiating surface will be too large; i.e., $R_{\xi} > R$, where $R_{\xi}$ is the source radius deduced using the spectral flux in equation (\ref{eq5}). An estimate of this artificial increase of the source size can be obtained by considering the frequency $\nu_{\rm p}$, where the spectral distribution in equation (\ref{eq5}) peaks. Writing $\hat{\nu}_{\xi} = \nu_{\rm p}/\nu_{\rm abs}$, one finds $\hat{\nu}_{\xi}^{-(p+4)\xi/2} = a(p)$. Since $a(p)$ is a function of $p$ only, $\hat{\nu}_{\xi} = \hat{\nu}_{\rm 1}^{1/\xi}$, where $\hat{\nu}_{\rm 1}$ is the normalized, standard synchrotron peak frequency, i.e., corresponding to $\xi = 1$. Hence, for a given $B$-value, the ratio of the peak spectral flux in equation (\ref{eq5}) to that for standard synchrotron radiation can be written
\begin{equation}
	\frac{f_{\xi}(\hat{\nu}_{\xi})}{f(\hat{\nu}_{\rm 1})} = \left[\hat{\nu}_{\rm 1}^{5/2}\left(1-\exp(-\hat{\nu}_{\rm 1}^{-(p+4)/2})\right)\right]^{1/\xi -1}.
	\label{eq6}
 \end{equation}
 
Approximating the ratio of the corresponding radiating surfaces by equating the energy radiated around the peak frequencies gives
\begin{equation} 
	\frac{R_{\xi}}{R_{\rm E}} \approx \left[\frac{\hat{\nu}_{\rm 1} f(\hat{\nu}_{\rm 1})}{\hat{\nu}_{\xi} f_{\xi}(\hat{\nu}_{\xi})}\right]^{1/2},
	\label{eq7}
\end{equation}
where $R_{\rm E}$ is the radius of a standard homogeneous source emitting the same amount of energy as the $\xi$-parameterization of the emissivity in equation (\ref{eq5}).
One can also compare the surfaces needed to give rise to the same spectral flux at the observed peak frequency (cf. the discussion above for SN 2011dh). Since $\hat{\nu}_{\xi} > \hat{\nu}_{1}$, a larger magnetic field strength is needed for the standard synchrotron emissivity than in the $\xi$-parameterization. Assuming $\nu_{\rm abs} \propto B$, the local emissivity at peak frequency scales as $B^2$, which yields
\begin{equation}
	\frac{R_{\xi}}{R_{\nu_{\rm p}}} \approx \left[\frac{\hat{\nu}_{\xi}^2}{\hat{\nu}_{\rm 1}^2} \frac{f(\hat{\nu}_{\rm 1})}{f_{\xi}(\hat{\nu}
	_{\xi})}\right]^{1/2},
	\label{eq7a}
\end{equation}
where, now, $R_{\nu_{\rm p}}$ is the radius of a standard source giving rise to the same spectral peak flux as that for the $\xi$-parameterization. Although observations indicate that the value of $p$ varies somewhat, $p=3$ is often used to fit the data. For this value of $p$, $\hat{\nu}_{\rm 1} \approx 1.1$ and, hence, $\hat{\nu}_{\rm 1} f(\hat{\nu}_{\rm 1})/\hat{\nu}_{\xi} f_{\xi}(\hat{\nu}_{\xi}) \approx 1.4^{(1/\xi -1)}$ and $\hat{\nu}_{\xi}^2 f(\hat{\nu}_{\rm 1})/\hat{\nu}_{\rm 1}^2 f_{\xi}(\hat{\nu}_{\xi}) \approx 2.0^{(1/\xi -1)}$. The value $\xi \approx 0.20$, derived for PTF11qcj by \cite{cor14}, gives $R_{\xi}/R_{\rm E} \approx 1.8$ and $R_{\xi}/R_{\nu_{\rm p}} \approx3.9$, respectively. 

The expansion velocity deduced by \cite{cor14} ($R_{\xi}/t \approx 0.3\,c$) places PTF11qcj  among the high velocity supernovae. However, as equations (\ref{eq7}) and (\ref{eq7a}) show, the actual  radius is sensitive to the details of the modeling using the parameter $\xi$. The larger value obtained for $R_{\xi}$ from equation (\ref{eq7a}) is likely to apply, since the fitting procedure was limited to the spectral region around the peak frequency. Hence, without a physical underpinning for $\xi$, it is possible to argue that the actual source radius has a value which may differ substantially from $R_{\xi}$.

\section{Inhomogeneities}\label{sect3}

The modelings discussed in section\,\ref{sect2} all assume a planar geometry, which, in principle, is inconsistent with the spherical geometry of the standard model. However, the range of optical depths in a homogeneous, spherical source is rather small \citep[see, for example,][]{f/b98} and does not seriously affect the conclusions. Therefore, in the discussion below of inhomogeneities and their observational consequences, the assumption of planar geometry is retained. A more detailed modeling of the observations will be done in a forthcoming paper, in which also the effects of a spherical geometry are considered.

In general, an inhomogeneous emission structure is caused by variations in the distribution of relativistic electrons and/or the magnetic field strength within the synchrotron source. As regards the effects on the emitted spectrum, the properties of an inhomogeneous magnetic field structure can qualitatively be divided into a few different types. When the average magnetic field strength varies over the projected source surface, its characteristics can, to a first approximation, be described by a source covering factor, $f_{\rm B,cov}$. Magnetic field structures with no variations over the projected source surface, can, in turn, be divided into two main types depending on their large scale properties. Either the magnetic field structure may be dominated by global variations (i.e., variations with depth within the source) or the magnetic field inhomogeneities may come primarily from local, small scale variations without any large scale structure. The first of these latter two situations was discussed in \cite{bjo13}. It was concluded that such magnetic field geometries are unlikely to give rise to flat-topped spectra/light curves. The second situation corresponds basically to a homogeneous source with a local synchrotron emissivity whose frequency distribution is broadened by a range of magnetic field strengths. Since the synchrotron frequency is proportional to $\gamma^2 B$, in order for this to cause a significant flattening of the emitted spectrum, the range of $B$-values must at least be as large as the square of the range of $\gamma$-values in the electron energy distribution. Although such a situation cannot be excluded, the large range of $\gamma$-values expected in shock acceleration makes this scenario less likely.

\subsection{The source covering factor}\label{sect3aa}

As suggested by the discussion above, if the flat-topped spectra/light curves are caused by an inhomogeneous emission structure, the qualitative properties of the inhomogeneities are quite restricted; for example, the magnetic field geometry should be such that it gives rise to a source covering factor $f_{\rm B,cov}$. Although the locally emitted spectrum would then still be the standard synchrotron one, $f_{\rm B,cov}$ would give rise to a range of optical depths over the source and, hence, broaden the observed spectrum. In this case, a direct relation between the properties of $f_{\rm B,cov}$ and $\xi$ is expected. In the analysis below, the source covering factor will be parameterized as $P(B) \propto B^{-a}$, where $P(B)$ is the probability to find a B-value between $B$ and $B+dB$. As a result, the covering factor can be written, $f_{\rm B,cov} \approx f_{\rm B_{\rm o},cov}\,(B/B_{\rm o})^{1-a}$ for $B_{\rm o}<B<B_{\rm 1}$. The observed spectrum then consists of three parts: A standard optically thick part for $\nu < \nu_{\rm abs}(B_{\rm o})$ ($F(\nu) \propto \nu^{5/2}$) and a standard optically thin part for $\nu > \nu_{\rm abs}(B_{\rm 1})$ ($F(\nu) \propto \nu^{-(p-1)/2}$). In between these two, there is a transition region in which the observed spectral flux is given by
\begin{equation}
	F(\nu) \propto \frac{R^2 \nu^{5/2} f_{\rm B,cov}}{B^{1/2}}.
	\label{eq8}
\end{equation}  
The frequency $\nu$ is related to the magnetic field strength $B$ through (see Appendix)
\begin{equation}
	\nu^3 \approx \nu_{\rm abs}^3 (B) \propto U_{\rm e} U_{\rm B}\,r \left(\frac{\gamma_{\rm min}}{\gamma}\right)^{p-2}.
	\label{eq9}
\end{equation}
It is possible that also the relativistic electrons are inhomogeneously distributed. If there exist a correlation with the distribution of magnetic field strengths, this will affect the spectral emissivity. Such an effect can be incorporated by introducing a third parameter $\delta$, defined by $U_{\rm e}\,r \,\gamma_{\rm min}^{p-2} \propto B^{\delta}$. One then finds
\begin{equation}
	\nu_{\rm abs}(B) =  \nu_{\rm abs}(B_{\rm o})\times (B/B_{\rm o})^{(p+2(1+\delta))/(p+4)}.
	\label{eq9a}
\end{equation}
Hence, an approximate expression for the spectral flux in the transition region is
\begin{equation}
	F(\nu) \approx F(\nu_{\rm abs}(B_{\rm o}))\left[\frac{\nu}{\nu_{\rm abs}(B_{\rm o})}\right]^{\frac{3p+7+5\delta -a(p+4)}{(p+2(1+\delta))}}.
	\label{eq10}
\end{equation}

In order for the low frequency part of the spectrum to be dominated by $B_{\rm o}$, the spectral index in equation (\ref{eq10}) should be smaller than $5/2$. It is seen directly from equation (\ref{eq8}) that this implies $a > 1/2$. Likewise, the high frequency part of the spectrum is dominated by $B_{\rm 1}$ when the spectral index in equation (\ref{eq10}) is larger than $-(p-1)/2$, which results in $a < (p+3)/2 + \delta$.  For $a$-values outside this range, the observed spectrum is that for a homogeneous source with a magnetic field given by $B_{\rm 1}$ ($a<1/2$) and $B_{\rm o}$ ($a > (p+3)/2 + \delta$), respectively. Another relevant value for $a$ is that giving a flat spectrum in the transition region. This corresponds to $a = a_{\rm o} \equiv (3p+7+5\delta)/(p+4)$.

This description of inhomogeneities introduces four new parameters ($f_{\rm B_{\rm o},cov}, a, B_{\rm 1}/B_{\rm o},$ and $\delta$). The observed values of $F(\nu_{\rm abs}(B_{\rm o}))$ and $ \nu_{\rm abs}(B_{\rm o})$ can no longer be used in equation (\ref{eq3}) to directly deduce values of $B_{\rm o}$ and $R$, since $F(\nu_{\rm abs})$ needs to be substituted by $F(\nu_{\rm abs}(B_{\rm o}))/f_{\rm B_{\rm o},cov}$. As will be discussed in section \ref{sect3b}, the value of $f_{\rm B_{\rm o},cov}$ can be constrained by the brightness temperature obtained from spatially resolved observations.

The spectrum in the transition region is degenerate for a spatially unresolved source, since the two observables depend on three model parameters; for example, both $\delta$ and $B_{\rm 1}/B_{\rm o}$ contribute to the spectral width of the transition region (a large value of $\delta$ can compensate for a small $B_{\rm 1}/B_{\rm o}$-value and vice versa) and the spectral slope in the transition region depends on both $a$ and $\delta$. This degeneracy can be broken by spatially resolved observations either quasi-simultaneously at two different frequencies within the transition region or at one given frequency at two different times. This would allow to determine the variation of the brightness temperature ($T_{\rm b} \propto (\nu_{\rm abs}(B)/B)^{1/2}$) with frequency or time, respectively, which is directly related to $\delta$ (cf. eqn (\ref{eq9a})). 

\subsection{The physical implications of the value for $\xi$}\label{sect3a}

As can be seen from figure \ref{fig1}, the value of $\xi$ affects the emitted spectrum in various ways; for example, the width of the transition from the optically thick to the optically thin part of the spectrum, the value of the peak frequency and the spectral flatness around the peak frequency. One should note that these variations of the spectral properties are correlated in the $\xi$-description. On the other hand, in the above discussion about inhomogeneities, they are each treated as independent aspects of the source. Hence, the value of $\xi$ deduced from observations is a compromise between reproducing these independent characteristics. However, even for an inhomogeneous source, the $\xi$-parameterization can be a useful method to characterize observed spectral deviations from the standard model; for exampel, the quality of observations only rarely warrant more than a one-parameter fit to such deviations and, as discussed above, for an unresolved source the parameters describing the inhomogeneities are degenerate.  Therefore, below, focus is on how a given $\xi$-value can be used to constrain the physical properties of an inhomogeneous source. 

\begin{figure}
\plotone{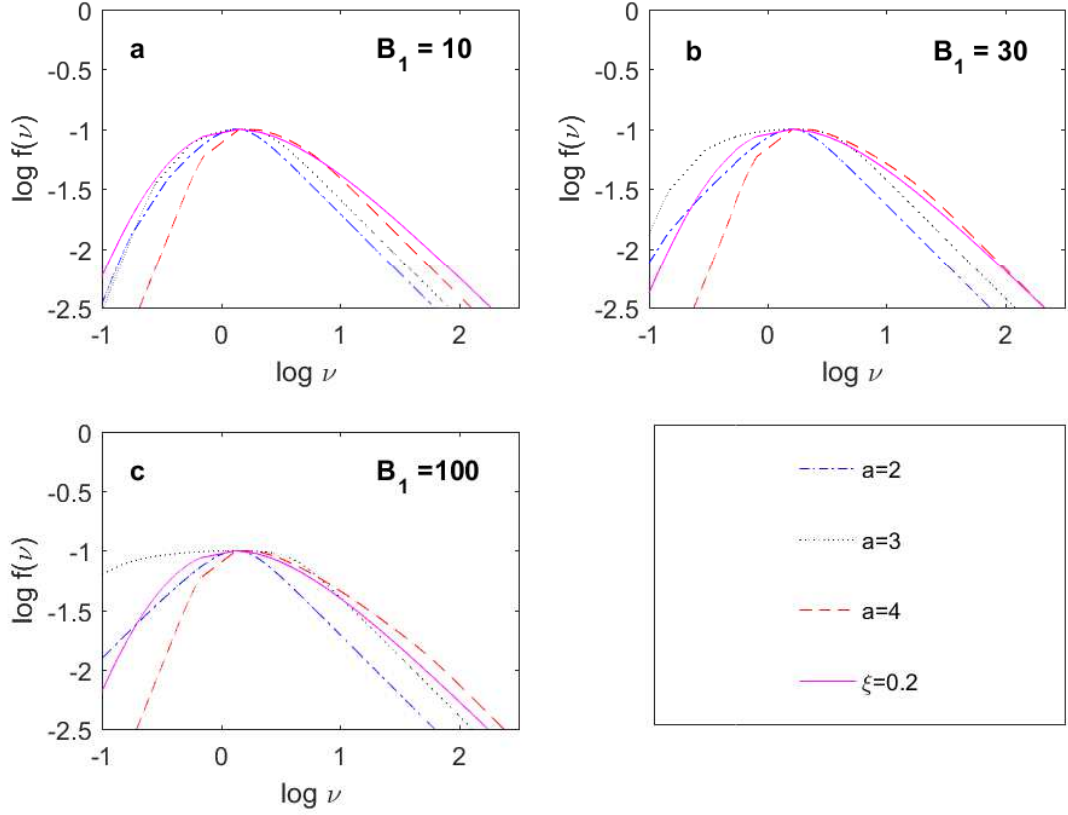}
\caption{Spectra from an inhomogeneous source for various values of $B_{\rm 1}$ and $a$ (cf. eqn (\ref{eq11})). All spectra have $\nu_{\rm abs} \propto B$. The spectra are normalized such that their peak frequencies as well as their peak spectral fluxes coincide. The modified local emissivity for $\xi = 0.2$ is shown for comparison.
\label{fig2}} 
\end{figure}

\begin{figure}
\plotone{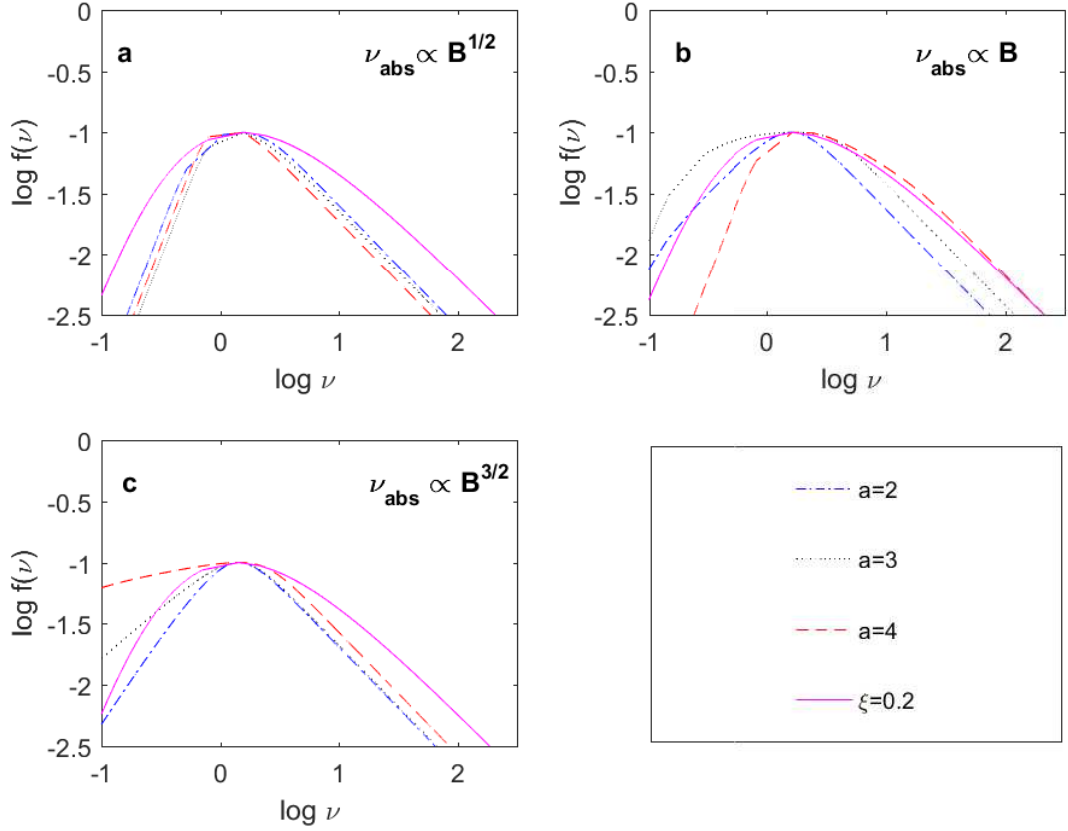}
\caption{Spectra from an inhomogeneous source for various relations between $\nu_{\rm abs}$ and $B$ (i.e., various values of $\delta$; see text) and values of $a$ (cf. eqn (\ref{eq11})). All spectra have $B_{\rm 1} = 30$. The spectra are normalized such that their peak frequencies as well as their peak spectral fluxes coincide. The modified local emissivity for $\xi = 0.2$ is shown for comparison. 
\label{fig3}} 
\end{figure}

The emitted spectrum from an inhomogeneous source is given by
\begin{equation}
	f(\nu) \propto \nu^{5/2}\int_1^{B_{\rm 1}} \frac{P(B)}{B^{1/2}}[1-\exp(-(\nu/\nu_{\rm abs}(B))^{-(p+4)/2})]dB.
	\label{eq11}
\end{equation}
For convenience, $B_{\rm o} = 1$ so that $B_{\rm 1}$ measures the range of B-values. Spectra are shown in figures \ref{fig2} and \ref{fig3} for  a representative set of values for $B_{\rm 1}$, $a$, and $\delta$. In order to compare the various spectra and how well they can be reproduced by a particular $\xi$-value, all spectra have been shifted so that both frequency and amplitude of the spectral peak coincide. Furthermore, instead of choosing specific values for $\delta$, the three different cases shown in figure \ref{fig3} correspond to different behavior of the brightness temperature ($T_{\rm b} \propto \nu^{(\delta -1)/(p+2(1+\delta))}$) in the transition region; namely, $\nu_{\rm abs} \propto B^{1/2}$ ($T_{\rm b}$ decreases with frequency), $\nu_{\rm abs} \propto B$ ($T_{\rm b}$ stays constant), and $\nu_{\rm abs} \propto B^{3/2}$ ($T_{\rm b}$ increases with frequency). Note that $\nu_{\rm abs} \propto B$ corresponds to $\delta = 1$ independent of the value of $p$.

\begin{figure}
\plotone{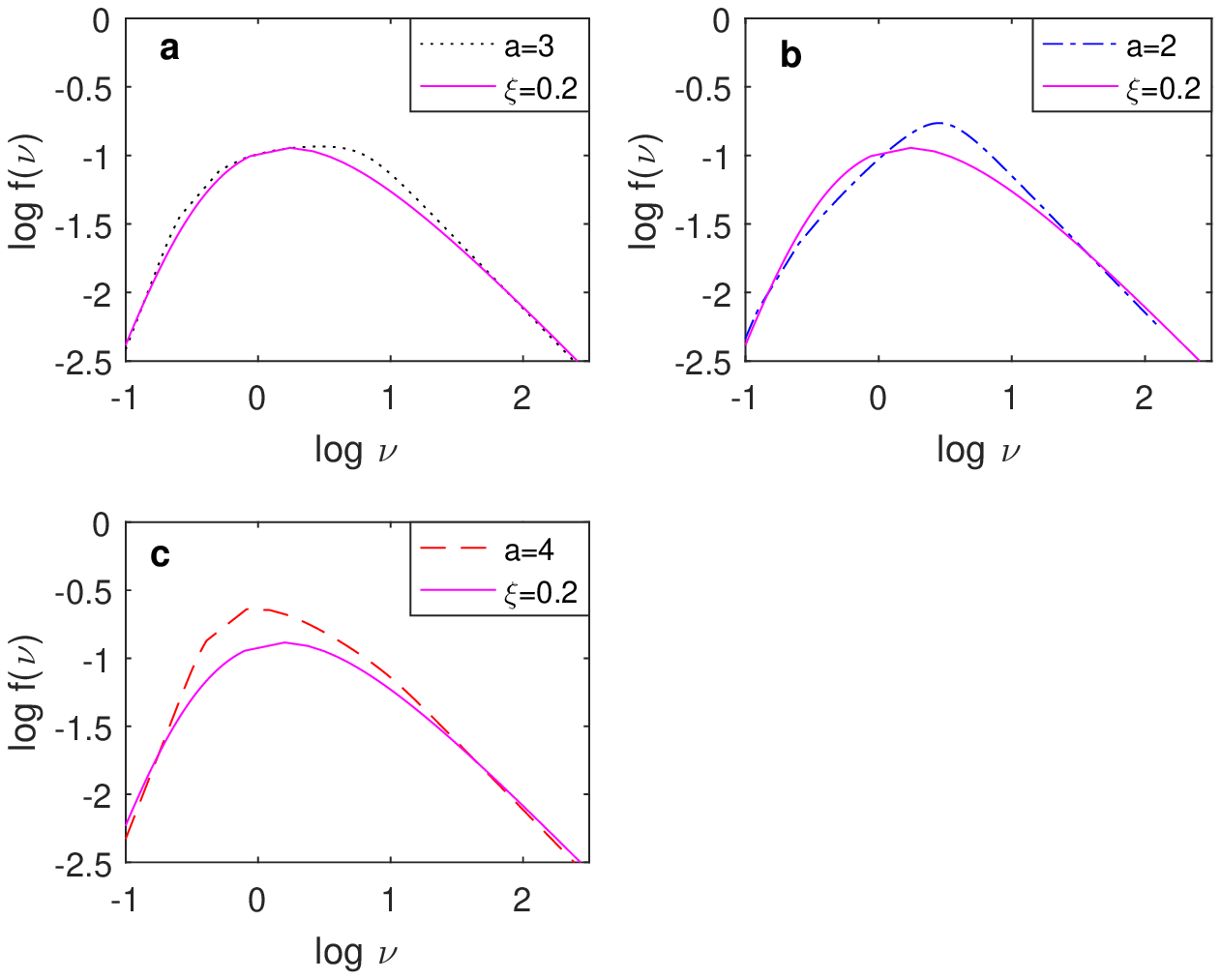}
\caption{Spectra from an inhomogeneous source with $B_{\rm 1} = 30$ and $\nu_{\rm abs} \propto B$ are shown individually for various values of $a$. In order to emphasize the importance of the spectral transition to the standard optically thick and thin spectral regions, spectra are normalized such that they overlap with the modified local emissivity ($\xi = 0.2$)  in the low and high frequency ranges, respectively.
\label{fig4}} 
\end{figure}

In figure \ref{fig4} three spectra are shown individually. Here, the fits are such that the spectral flux in the low and high frequency ranges overlap.  In addition, in figure \ref{fig4}a a spectral fit  is attempted also around the peak frequency. Although no detailed fitting has been made, it is seen that $\xi = 0.2$ implies $B_{\rm 1} \approx 30$ for $\delta = 1$ (i.e., $\nu_{\rm abs} \propto B$). Another thing to notice is that a "good" fit is obtained for $a \approx a_{\rm o}$. This corresponds to a flat spectrum, where the different $B$-values contribute approximately equally to the spectral flux. Other $a$-values give rise to more peaked spectra so that for  $a < a_{\rm o}$ the spectrum becomes increasingly dominated by the larger $B$-values, while for  $a > a_{\rm o}$ the smaller $B$-values dominate. 

It is clear that a $\xi$-parameterization of the inhomogeneous spectra in figures \ref{fig4}b and \ref{fig4}c would give a poor fit. However, it is seen that the asymmetry of the spectral distribution around the peak frequency is different in the two cases. It may, therefore, be possible to get a rough estimate of the $a$-value from the sign-change in the residuals around the peak frequency resulting from a $\xi$-parameterization of the observations. 

Another aspect of  figures \ref{fig4}b and \ref{fig4}c is that when observations are limited to the spectral range around the peak frequency, they would likely be fitted by a value of $\xi$ larger than $0.2$. As mentioned above, although the range of $B$-values may be large, it is the value of $a$ which determines the fraction of these $B$-values actually contributing significantly to the emitted radiation. Hence, using $a = a_{\rm o}$ and the value of $\xi$ obtained from fitting only the spectral distribution around the peak frequency would give a rough estimate of the range of relevant $B$-values in the inhomogeneous model. 

As discussed above, spatially resolved observations are needed to deduce individual values for $B_{\rm 1}$ and $\delta$. However, as equation (\ref{eq8}) shows, the variation of the covering factor with frequency depends only weekly on $\delta$; i.e., $f_{\nu, {\rm cov}} \propto F(\nu)\,\nu^{-2}\,T_{\rm b}^{-1}$. For the broad/flat spectra discussed here, it is seen that the maximum covering factor occurs at a frequency substantially below that of the spectral peak. Hence, detailed observations at low frequencies are needed to determine the range of covering factors and, in particular, its maximum value. 

In order to illustrate how inhomogeneities can affect the deduced source radius consider again the spectral distribution from the $\xi$-parameterization. Assume, for simplicity, $\delta = 1$ so that $T_{\rm b} = $\,const. The emitting area at maximum covering factor is then obtained directly from equation (\ref{eq7a}) with $\hat \nu_{\xi}$ replaced by $\hat \nu_{\rm min}$, where $\hat \nu_{\rm min}$ is the minimum frequency for which a homogeneous source can give substantial contribution to the emitted spectral flux. The value of $\hat \nu_{\rm min}$ can be estimated as follows: Since $f(\hat \nu_{\rm 1}) \propto \hat \nu_{\rm 1}^{5/2}(1 - \exp (-\hat \nu_{\rm 1}^{-(p+4)/2})) = 0.65$ for $p = 3$ (cf. eqn (\ref{eq6})), the frequency on the extrapolated $\nu^{5/2}$ part of the spectrum with the same value is then $\hat \nu_{\rm ex} = 0.65^{2/5} = 0.84$. The value of $\hat \nu_{\rm min}$ is the frequency for which $f_{\xi}(\hat \nu_{\rm min})$ both has the same value as the extrapolated $\nu^{5/2}$ part and is separated from it in frequency by a factor $\hat \nu_{\rm 1}/\hat \nu_{\rm ex}$. This leads to
\begin{equation}
	\left(\hat \nu_{\rm min}\, \frac{\hat \nu_{\rm ex}}{\hat \nu_{\rm 1}} \right)^{5/2} = \hat \nu_{\rm min}^{5/2} \left(1 - \exp (-\hat \nu_{\rm min}^{-7\xi/		2})\right)^{1/\xi}.
	\label{eq12}
\end{equation}
With $\xi = 0.2$, this yields $\hat \nu_{\rm min} = 0.38$. Equation (\ref{eq7a}) then gives $R_{\xi} / R_{\nu_{\rm min}} = 1.3$, where $R_{\nu_{\rm min}}$ is the radius of the emitting area at maximum covering factor. An alternative way to obtain the same result is to note that $\hat \nu_{\rm min}$ is obtained by requiring the spectra from the homogeneous source and the $\xi$-parameterization to overlap in the standard optically thick part. Since radius scales as $B^{1/4}$, $R_{\xi} / R_{\nu_{\rm min}} = (\hat \nu_{\rm 1}/\hat \nu_{\rm min})^{1/4} = 1.3$. Furthermore, it is seen from equation (\ref{eq7a}) that $R_{\nu_{\rm min}} \approx 3R_{\nu_{\rm p}}$, so that interpreted as inhomogeneities,  $\xi = 0.2$ corresponds to a covering factor at $\nu_{\rm p}$ smaller by a factor $\approx 9$ than its maximum value at lower frequencies. 

This shows that for $f_{\rm B_{o}, cov} \approx 1$ the radius of an inhomogeneous source can be smaller than the one deduced from a $\xi$-parameterization of the observed radiation.

\subsection{The brightness temperature of SN 1993J}\label{sect3b}

The spatially resolved VLBI-observations of SN 1993J are still those with the highest quality of any radio supernova. They showed that the average observed brightness temperature of the source ($T_{\rm b}^{\rm obs}$) was rather low.  The evolution of the radio emission has been modeled assuming the validity of the standard model, i.e., $f_{\rm B,cov}=1$ \citep{f/b98,per01}.  The low value of $T_{\rm b}^{\rm obs}$ was accounted for by low values of $U_{\rm e}/U_{\rm B}$ as well as $r/R$ (cf., eqns (\ref{eq3}) and (\ref{eq4})); the latter caused by synchrotron cooling due to the large $B$-value. 

Although the standard model gives a good fit, such large $B$-values strain the consistency of the model; for example, the implied mass/energy associated with the high velocity ejecta is more than an order of magnitude larger than thought possible for standard supernovae. An alternative is to assume $B$-values small enough for synchrotron cooling to be unimportant and attribute the low value of $T_{\rm b}^{\rm obs}$ to a covering factor below unity \citep[see][for details]{bjo15}. Since observations indicate a small range of $B$-values, $T_{\rm b}^{\rm obs} = f_{\rm B,cov} T_{\rm b}$. 

During the later phases when $F(\nu_{\rm abs})$ is roughly  constant, the value of $T_{\rm b}^{\rm obs}$ is lower than its equipartition value (i.e., $U_{\rm B} = U_{\rm e}$) by a factor $2$-$3$. If this, instead, is attributed to inhomogeneities, $f_{\rm B,cov} \approx 1/3 - 1/2$ is implied for equipartition. Since the optically thin spectral index stays constant, cooling must be unimportant during the whole observing period. This gives a magnetic field strength lower by, at least, a factor $\approx 8$ than used in \cite{f/b98}. Since, for a given value of $T_{\rm b}^{\rm obs}$, $f_{\rm B,cov} \propto B^{1/2}$, this shows that conditions in SN 1993J are consistent with equipartition (although $U_{\rm e}>U_{\rm B}$ cannot be excluded). 

Additional support for attributing the low values of $T_{\rm b}^{\rm obs}$ in SN 1993J to inhomogeneities comes from the observations of \cite{bie03} and \cite{bie08}. They emphasize the presence of time-dependent brightness modulation within the circular outer contours. Similar structures are seen also in SN 2011dh. However, as discussed in \cite{dew16}, in contrast to SN 1993J, it is possible that they result from the data reduction procedure. The VLBI-observations of SN 1993J thus show that even when no broadening of the spectra/light curves is apparent, the presence of inhomogeneities cannot be excluded.

There is also a third possibility to explain the low value of $T_{\rm b}^{\rm obs}$ in SN 1993J. When the free-free absorption at $\nu_{\rm abs}$ is larger than unity (i.e., $\tau_{\rm ff}(\nu_{\rm abs}) > 1$), the observed peak in the synchrotron spectrum is not due to self-absorption and, hence, results in a lower brightness temperature \citep[see, e.g.,][]{che82b}. However, there are several reasons why this is an unlikely explanation for the low value of $T_{\rm b}^{\rm obs}$. After about a few hundred days, no measurable effects of free-free absorptions are apparent. As mentioned above, in spite of this, the value of $T_{\rm b}^{\rm obs}$ is lower by a factor $2$-$3$ as compared to its equipartition value. Furthermore, although modeling indicates that free-free absorption is present during the earlier phase, its main effect is to steepen the optically thick part of the synchrotron spectrum; for example, $\tau_{\rm ff}(\nu_{\rm abs}) < 1$ even at the earliest observations at $\approx 11$\,days.

\subsection{Dips in the light curves}\label{sect3c}

It is quite common for radio light curves to show dips. They are usually attributed to a varying mass-loss rate from the progenitor star; for example, SN 2001ig \citep{ryd04} and SN 2011dh \citep{dew16}. As pointed out by \cite{wei07}, also SN 1993J has a dip in the  radio light curve around $\sim 460$\,days for the shortest observed wavelength (1.2\,cm). However, there are two aspects of this dip that suggest it may not entirely be caused by deviations from the standard radial density distribution of the circumstellar medium. The dip is not apparent in the other optically thin light curves at longer wavelengths (at this time, synchrotron self-absorption sets in at $\sim$20\,cm). Hence, for example, it is not due to a sudden increase in the decline of $\nu_{\rm abs}$, since this would affect all the optically thin frequencies alike. One way to account for the observed behavior is to assume the existence of optically thick inhomogeneities in this spectral range, which, in the standard model, would correspond to optically thin emission. In the most extreme case, this emission could be optically thick, inhomogeneous emission. The observed spectral index together with the approximation in equation (\ref{eq10}) then implies $a \approx 2.9$ with $p=3$ and $\delta = 0$. The rapidly decreasing value of $\nu_{\rm abs}$ in SN 1994I (see sect. \ref{sect2aa}) could have a similar origin; i.e., a decreasing value of $B_{\rm 1}/B_{\rm o}$ but without any change in the evolution of $B_{\rm o}$.

The well observed X-ray light curve in SN 1993J has a dip coincident with and very similar to that at 1.2\,cm \citep{wei07}. Since the X-ray emission is thought to come from the shocked ejecta material, the inhomogeneous radio emission is then likely to be associated with the reverse shock as well. It was suggested in \cite{bjo15} that the simultaneous, achromatic breaks in all the radio light curves as well as those in the X-ray range at $\sim 3100$\,days were due to a flattening in the density distribution of the ejecta. The energy/momentum input from the ejecta would then not be sufficient to maintain the self-similar shock structure, which would lead to a weaker reverse shock. A similar cause is possible for the dips at $\sim 460$\,days. If the amplification of the magnetic field is due to turbulence driven by the Rayleigh-Taylor instability at the contact discontinuity \citep{che92}, a weaker reverse shock could narrow down the range of magnetic field strengths, i.e., a smaller value of $B_{\rm 1}/B_{\rm o}$. In this scenario, the dips would be caused by the ejecta structure rather than that of the circumstellar medium. The reason for the transient weakening of the reverse shock is not clear but one may note that it occurs when the optical nebular lines fade away and H$\alpha$, as well as other lines, acquires a box-like profile. As emphasized by \cite{mat00}, this indicates a transition to a new emission phase dominated by the effects of circumstellar interaction.

\section{Discussion}\label{sect4}

There is increasing evidence that a standard homogeneous source model does not capture all the main characteristics of the observations. It was shown in section \ref{sect3} how many of these non-standard features can be accounted for by inhomogeneities within the source. Since most observations of radio supernovae are spatially unresolved, a central issue is how this affects the radius of the source. It was concluded that the deduced value of the radius is quite sensitive to the details of the modeling; in particular, the transition at low frequencies to the standard optically thick part of the synchrotron spectrum. Therefore, when there are indications of deviations from the standard model, the observationally derived radius should be treated with care.

It is clear that fitting a standard homogeneous model to observations around the peak frequency gives a lower limit to the source radius. Sometimes a parameter $\xi$ is introduced to artificially flatten the intrinsic spectrum around the peak frequency. This, on the other hand, tends to give a radius larger than that resulting from an inhomogeneous model, at least when the total covering factor is close to unity. 

In spatially unresolved sources, the velocity of the forward shock is usually derived from the increasing source radius. In a few radio supernovae, the deduced shock velocity is so large that it strains the standard model. It has been suggested \citep{sod10} that these large velocities are due to an additional input of momentum/energy from a central engine. Hence, it is important to establish the possible occurrence and effects of inhomogeneities before deciding whether the presence of a central engine is indicated by the observations. Furthermore, in standard supernovae, the decline of the shock velocity with time depends on the density structure of the ejecta. As discussed for SN 2011dh, the conclusions drawn from combining spatially unresolved observations during the early phase of the evolution with spatially resolved VLBI-observations during the later phase can be seriously affected by the presence of source inhomogeneities.

\subsection{Relation to supernova remnants}\label{sect4aa}

The large sizes of supernova remnants (SNR) allow a much more detailed study of their spatial structure than is possible for SNe in their earlier phases of evolution. There are a number of well-observed SNRs in which the ejecta input of momentum/energy is large enough for a distinct reverse shock to form. Such SNRs may throw light on the physical processes shaping the region in between the forward and reverse shocks. One characteristic aspect of SNRs is that their spatial structure both in radio and X-rays appears to be the result of two different processes; namely, the presence of thin outer rims together with an inner extended ring of emission; for example, Tycho/SN 1572 \citep{dic91} and Cassiopeia A \citep{got01}. The inner ring of radio emission is well correlated with the distribution of thermal X-ray emission, while the outer thin rims seem to be devoid of thermal emission so that both the radio and X-rays are dominated by synchrotron radiation \citep{tra15,got01}. Furthermore, since the rims in these two frequency ranges are roughly co-spatial, the X-ray part is likely the high frequency extension of the radio emission. 

\cite{j/n96a,j/n96b} argued that the inner and broader component in Tycho/SN 1572 is due to a magnetic field amplified by the turbulence driven by the Rayleigh-Taylor instability emanating from the contact discontinuity. They also pointed out that the outer thin rim is not the result of just compression of the magnetic field by the forward shock, since polarization shows that the magnetic field has a dominant radial component \citep{dic91}. 

Although it is likely that the width of the X-ray rims is affected by synchrotron cooling \citep{res14}, this is not so for the radio rims. Hence, observations indicate a strong magnetic field limited to a narrow region behind the forward shock. MHD calculations are unlikely to account for such a structure \citep{guo12}. The main problem is the rather long growth time for the MHD-instabilities, which is set by the eddy turn-around time. As a result, the advection of the fluid downstream causes the magnetic field strength to increase with distance behind the shock. Simulations show that neither the distribution of magnetic field behind the shock nor its strength relative the magnetic field amplified by the Rayleigh-Taylor instability is consistent with observations. 

On the other hand, the association of the inner radio ring with the turbulent entrainment of the ejecta into the region with shocked circumstellar gas is supported by its rather close correlation with the thermal X-ray emission. \cite{che92} showed that the entrained ejecta gas does not quite reach the forward shock. Since the thin rims show little evidence for thermal X-ray emission, they would then be associated with the narrow region behind the forward shock not affected by the turbulence driven by the Rayleigh-Taylor instability and dominated by the high temperature gas from the shocked circumstellar medium.

There is a group of SNRs with bilateral symmetry in the radio and medium/hard X-ray regimes, with SN 1006 as its most prominent member \citep[see][for a review]{kat17}. It is thought that this barrel-like symmetry is caused by the supernova exploding in a region with a coherent large scale magnetic field. Support for such an origin comes from radio polarization mesurements \citep{rey13}, which show the magnetic field to be radial in the regions with the most intense synchrotron radiation but tangential in the perpendicular direction where the emission is much weaker; i.e., the magnetic field direction is roughly constant inside the forward shock and, presumably, corresponds to that in the ambient medium. This differs from SNRs with circular symmetry in which the magnetic field direction is preferentially radial in the whole region interior to the forward shock. 

In contrast to the radio and medium/hard X-rays, the soft X-ray and H$\alpha$ emission in SN 1006 are roughly spherically distributed, which suggests a spherically outflowing ejecta and that the bilateral symmetry is primarily restricted to the non-thermal emission. Since the Rayleigh-Taylor instability is driven by the momentum/energy input at the reverse shock, the resulting magnetic field should also be spherically distributed. Hence, the extent of the non-thermal emission would then be determined by the distribution of relativistic electrons rather than the magnetic field. This supports acceleration of particles through the first order Fermi process at the shock front rather than, for example, second order Fermi acceleration associated with the Rayleigh-Taylor driven turbulence. The latter is likely to result in a correlation between magnetic field strength and density of relativistic electrons, which is not observed.

\subsection{Inhomogeneities and shock physics}\label{sect4a}

One aspect of the inhomogeneities is their possible relation to the details of the shock structure and, hence, the micro physics governing the formation and evolution of shocks. During recent years, detailed numerical modeling from first principles (PIC-simulations) has become possible of the physical environment of a shock. This includes the amplification of the magnetic field strength and the injection as well as acceleration of relativistic particles \citep{c/s14c,cap15,par15}. Although such calculations are still rather limited in range, there are several emerging properties which may well be generic to the shock-induced processes. Some of these have a direct bearing on the inhomogeneous model discussed above. As the quality of the radio data has likewise increased substantially, this makes it timely to asses the synergies that can be gained from a comparison of the results from theses two areas.

The amplification of the magnetic field strength is sensitive to the direction of the background field in the upstream region. For very oblique shocks (magnetic field aligned mainly with the shock front), the protons/ions never enter the diffusive shock acceleration process. As a result, the magnetic field experiences little amplification except that due to compression by the shock. Efficient amplification of the magnetic field and injection/acceleration of particles occur instead for quasi-parallel shocks. An important feature of such flows is the occurrence of a filamentary instability, which excites modes transverse to the magnetic field. The characteristics of the ensuing filamentary structure of the magnetic field impact the parameter values of the inhomogeneous model in several ways. 

Due to the direction of the background field, the filamentary structure of the strongest magnetic fields is aligned roughly parallel to the shock normal. Therefore, in addition to variations of  the magnetic fields strength with distance from the shock front, there is also a substantial variation over the shock surface of its average projected strength. Such a situation can, to a first approximation, be described by a source covering factor. This is in line with the inhomogeneous model discussed in section\,\ref{sect3}. Furthermore, the filaments of strong magnetic fields enclose under-dense regions of thermal plasma. These cavities are instead filled with particles energized by the acceleration process. This indicates that a certain amount of anti-correlation between magnetic field strength and energy density of relativistic electrons is to be expected. In the inhomogeneous model, this translates to $\delta<0$ (cf. eqn (\ref{eq9a})) and, hence, for a given observed spectral broadening such an anti-correlation would increase the underlying spread in magnetic field strengths, i.e., a larger value of $B_{\rm 1}$.

Within the framework of the PIC-simulations, the main parameter affecting the result is the Alfv\'{e}nic Mach number ($M_{\rm A}$) of the shock; for example, the mean value of the amplification of the magnetic field scales roughly as $M_{\rm A}^{1/2}$. Since the spread in $B$-values scales in a similar manner, the value for $B_{\rm 1}$ obtained from observations is directly related to $M_{\rm A}$. Although detailed modeling is needed in order to characterize the magnetic field structure, for the most extreme radio supernova (PTF11qcj), $B_{\rm 1} \sim 30$ was indicated assuming the spatial distribution of the magnetic field strength and the density of relativistic electrons to be correlated (i.e., $\delta = 1$). Since PIC-simulations suggest $\delta < 0$, this is likely a lower limit for $B_{\rm 1}$ (cf. eqn (\ref{eq9a})). Furthermore, as discussed in section\,\ref{sect3a}, the magnetic field distribution may be such that the low and/or high end do not contribute substantially to the emitted radiation, which would further increase the actual spread in $B$-values. This, then, implies $M_{\rm A} \gsim 10^2$.

With such a high value for $M_{\rm A}$, PIC-simulations indicate that the filamentary magnetic field structure close to the shock should transition into a turbulent state not far behind the shock. As discussed in section\,\ref{sect3}, synchrotron emission from a highly turbulent (i.e., isotropic) magnetic field is unlikely to result in an appreciable broadening of the spectrum/light curve. However, the magnitude of the magnetic field amplified at the shock is expected to decrease with distance behind the shock. This is seen in the simulations by \cite{c/s14b}. Hence, it is possible that the emitted synchrotron radiation comes mainly from a narrow region around the shock, in which the relevant magnetic field structure would be dominated by the filamentary instability associated with the shock.

Such a situation would imply that only a small fraction of the region of shocked gas contributes to the radio emission. Since, on the other hand, the accelerated electrons should fill up the whole of this region, this would affect the expected amount of inverse Compton scattered radiation coming from the source. \cite{bjo13} showed that a good correlation between the radio, optical and X-ray luminosities exists for Type Ib/c SNe, including those suggested to be powered by a central engine. It was argued that this was best explained by a model in which the X-ray emission was inverse Compton scattering of the optical emission by the same electrons giving rise to the radio emission. However, to be consisted, observations required a low filling factor for the radio emission; for example, equipartition between relativistic electrons and magnetic fields indicated a filling factor $\sim 10^{-3} - 10^{-2}$. 

The need for a roughly radial field for efficient injection of thermal particles into the acceleration process in PIC-simulations is consistent with the observed properties of the rims in SNRs. The width of the rims may then be determined by the shock formation process. Although PIC-simulations show a magnetic field strength decreasing with distance behind the shock, the radial extent of the simulations is much smaller than the observed rim widths. It is not clear how the length scale for this decrease varies with the available computational spatial extent, since, for example, this also determines the maximum momentum of the particles reached in the simulations. 

The observed projected rim widths in SNRs indicate that their radial extent is at most a few percent of the radius. With an assumption of a basic qualitative similarity between SNRs and the earlier phases of SNe, the low filling factor in Type Ib/c SNe may be accounted for by associating their radio emission with the rims in SNRs. The main difference between SNRs and Type Ib/c SNe would then be the relative strength of the emission from the outer rims and the inner ring; in the former, ring emission dominates while in the latter radio emission comes primarily from the thin rims.

\subsection{Inhomogeneities and the Rayleigh-Taylor instability}\label{sect4b}

The alternative origin for the magnetic field amplification is the turbulence generated by the Rayleigh-Taylor instability at the contact discontinuity. As discussed by \cite{j/n96a,j/n96b}, the strongest magnetic fields are found around the high density fingers protruding radially from the contact discontinuity. Just like the filamentary instability around the shock, this gives rise to a magnetic field structure, which, to a first approximation, can be described by a source covering factor. The region affected by the Rayleigh-Taylor instability covers a fair fraction of the total volume between the forward and reverse shocks, while the filamentary instability is expected to be limited to the region around the forward shock. Hence, although a detailed comparison is not yet possible, mainly due to the limit scope of the PIC-simulations, it does suggest a substantially larger volume filling factor of the magnetic field to result from the Rayleigh-Taylor instability than from the filamentary instability. This would be in line with the structural properties of SNRs, in that their thin outer rims would be due to processes described in the PIC-simulations while the extended inner ring results from the Rayleigh-Taylor instability.

Although no simulations have been made where both of these effects have been considered simultaneously, it is likely that one of these instabilities does not preclude the presence of the other. Hence, their relative strength depends on the driving mechanisms. While the physics of the forward shock is determined to a large extent by $M_{\rm A}$, the Rayleigh-Taylor instability is very sensitive to the deceleration of the supernova ejecta and, hence, the value of $n$. Since these two factors are not likely to be closely correlated and, furthermore, are expected to vary at least from one type of supernova to another, this would result in a range of observed properties. Therefore, high quality radio observations may prove important to elucidate not only the relative importance of these two processes but also their relation to other observed properties of SNe.

The amplification of the magnetic field at the shock front is intimately connected to the acceleration of particles. This is in contrast to the magnetic field amplification due to the Rayleigh-Taylor instability. As suggested by the observed properties of SNRs, the relativistic electrons in both scenarios would be produced by first order Fermi acceleration at the forward shock; hence, the observed differences of the radio emission should derive mainly from the properties of the magnetic field. There are then two characteristics that may be used to observationally distinguish between the two scenarios for the origin of the magnetic field. 

Since the processes amplifying the magnetic field are fundamentally different, it is likely that also the properties of the resulting inhomogeneities are different. However, at the present time, a detailed comparison between the two does not seem feasible. A more direct route is to compare the observed implications of the different sites and extent of the magnetic field structure posited by the two scenarios. As already discussed in section\,\ref{sect4a}, the observations give support for a narrow radio emission region in Type Ib/c SNe associated with the forward shock. 

On the other hand, \cite{bie11} \citep[see also][]{mar09} showed that the radio emission in SN 1993J (Type IIb) had a shell like structure, whose extent corresponded roughly to the expected region in between the shocks. It was argued in \cite{bjo15} that the radio emission in SN 1993J is best understood as coming from the Rayleigh-Taylor unstable region emanating from the contact discontinuity; in particular, the almost constant velocity observed during the first few hundred days (when the shell was not spatially resolved) was caused by a combination of a decelerating forward shock and an expansion of the Rayleigh-Taylor unstable region. Only after the instability had saturated did the observed velocity correspond to that of the forward shock. 

The indications of different regions for the origin of radio emission are in line with other observed properties. The deceleration of the forward shock in SN 1993J is stronger than typically observed in Type Ib/c SNe ($n\approx 7$ vs $n\approx 12$), causing a stronger driving of the Rayleigh-Taylor instability. Even though the radio emitting region in SN 1993J would be substantially larger than in Type Ib/c SNe, the observed X-ray emission from SN 1993J is actually larger than expected from the correlation between radio, optical, and X-ray luminosities in Type Ib/c SNe. This is due to the much higher density of the circumstellar medium in SN 1993J as compared to Type Ib/c SNe, which causes bremsstrahlung to dominate the inverse Compton scattering of the optical emission. Furthermore, in analogy with SNRs, one would expect the thermal X-rays from the shocked ejecta to be associated with the radio emission. As discussed in section\,\ref{sect3c}, the similar variations in radio and X-rays in SN 1993J support such a close relation.

\section{Conclusions}\label{sect5}

The main result of the present paper is that modeling radio supernovae as homogeneous, spherically symmetric synchrotron sources does not encompass several of their observed characteristics. It is shown that these deviations from the standard model can be accounted for by an inhomogeneous source structure. If inhomogeneities are important, a few general conclusions follow:

1) The radius of the source can differ substantially from that resulting from a homogeneous model. A few examples are discussed in sections \ref{sect2} and \ref{sect3} chosen in order to illustrate that the radius deduced from observations is quite sensitive to the details of the modeling.

2) The flat-topped spectra/light curves severely constrain the properties of the inhomogeneities; e.g., the magnetic field structure needs to have a dominant component of filaments in the radial direction, so that the average magnetic field strength varies over the source surface.

The properties of the inhomogeneities are likely determined by the same mechanism, which amplifies the magnetic field and, possibly, also accelerates the relativistic electrons. This opens up for a direct, detailed comparison between theoretical modeling and observations.

3) The filamentary structure implied for the magnetic field is consistent with an origin either from the turbulence driven by the Rayleigh-Taylor instability at the contact discontinuity or the filamentary instability expected to be associated with the forward shock. However, in the latter case, the magnetic field strength needs to decrease with distance behind the shock so that the radio emission is coming mainly from the narrow region where the filaments are most apparent.

4) Observations indicate that both of these mechanisms are important but in different types of supernovae. Many aspects of SN1993J, a Type IIb supernova, find an explanation with a radio emission region determined by the Rayleigh-Taylor instability. On the other hand, the observed properties of Type Ib/c SNe suggest a small filling factor for the radio emission, which would be consistent with a thin rim at the forward shock.

5) The cause for the variation of the relative strength between these mechanisms is not clear. However, the spatially resolved structure of supernova remnants is consistent with both of them being present simultaneously, although the magnetic field amplified by the Rayleigh-Taylor instability is the dominating one. Hence, supernova remnants seem to have retained some of the characteristics of their youth. \\[1.5cm]

\appendix

\begin{center}
{\bf Appendix}
\end{center}
\section{Synchrotron formulae}

The synchrotron formulae in the paper are written in a form useful for the discussion of various aspects of inhomogeneities. Since they are somewhat non-standard, their derivation is given below.

The local spectral synchrotron emissivity is \citep[e.g.,][]{r/l79}
\begin{equation}
	P(\nu) \propto B \int f(\frac{\nu}{\gamma^2 B}) N(\gamma){\rm d}\gamma,
	\label{eqa1}
\end{equation}
where $f$ is the normalized, single electron emission spectrum, which peaks at $\nu \propto \gamma^2 B$. For an electron energy distribution, $N(\gamma)$, varying more slowly than $f$ around $\gamma_{\nu} \propto (\nu/B)^{1/2}$, $P(\nu) \propto B \gamma_{\nu} N(\gamma_{\nu})$.
Likewise, the absorption coefficient is 
\begin{equation}
	\alpha(\nu) \propto \nu^{-2} \int f(\frac{\nu}{\gamma^2 B}) \frac{N(\gamma)}{\gamma}{\rm d}\gamma.
	\label{eqa2}
\end{equation}
Again, with a slowly varying $N(\gamma)$ at $\gamma_{\nu}$, $\alpha(\nu) \propto \nu^{-2} B N(\gamma_{\nu})$. Instead of $N(\gamma)$, it is sometimes convenient to use the energy density of electrons at $\gamma$, i.e., $U_{\rm e}(\gamma) = \gamma^2 N(\gamma) mc^2$, which yields
\begin{eqnarray}
	\alpha(\nu) & \propto &  B\,U_{\rm e}(\gamma_{\nu})/(\nu^{2}\gamma_{\nu}^2) \nonumber\\
	 & \propto &  \nu^{-3} U_{\rm B}U_{\rm e}(\gamma_{\nu}).
	\label{eqa3}
\end{eqnarray}
The optical depth can then be written
\begin{eqnarray}
	\tau(\nu) & = & \alpha(\nu)\,r \nonumber\\
	& \propto &  \frac{B^4}{\nu^3} \frac{U_{\rm e}(\gamma_{\nu})}{U_{\rm B}}\frac{r}{R}\,R.
	\label{eqa4}
\end{eqnarray}
Together with
\begin{equation}
	F(\nu) \propto \nu^{5/2} \frac{R^2}{B^{1/2}}(1-\exp{(-\tau(\nu))}),
	\label{eqa5}
\end{equation}
this can be used to solve for $\tau(\nu = \nu_{\rm abs}) = 1$ in the standard way,
\begin{eqnarray}
	BR & \propto & F(\nu_{\rm abs})^{6/17} y^{-5/17}\nonumber\\
	B & \propto & \nu_{\rm abs}F(\nu_{\rm abs})^{-2/17} y^{-4/17},
	\label{eqa6}
\end{eqnarray}
where
\begin{equation}
	y \propto \frac{U_{\rm e}(\gamma_{\nu_{\rm abs}})}{U_{\rm B}}\frac{r}{R}.
	\label{eqa7}
\end{equation}

When the electron energy distribution is a power-law, $N({\gamma}) \propto \gamma^{-p}$ for $\gamma > \gamma_{\rm min}$ and $p>2$, equations (\ref{eqa6}) and (\ref{eqa7}) can be rewritten as
\begin{eqnarray}
	BR & \propto & F(\nu_{\rm abs})^{\frac{p+4}{2p+13}} \hat{y}^{\frac{-5}{2p+13}}\nonumber\\
	B & \propto & \nu_{\rm abs}F(\nu_{\rm abs})^{\frac{-2}{2p+13}} \hat{y}^{\frac{-4}{2p+13}},
	\label{eqa8}
\end{eqnarray}
where
\begin{equation}
	\hat{y} \propto \frac{U_{\rm e}}{U_{\rm B}}\frac{r}{R}\gamma^{p-2}_{\rm min}.
	\label{eqa9}
\end{equation}
Here, $U_{\rm e} = U_{\rm e}(\gamma) \left(\frac{\gamma}{\gamma_{\rm min}}\right)^{p-2}$ is the total energy density of relativistic electrons.

\clearpage

\clearpage

\begin{thebibliography}{}
 
    \bibitem[Alexander et al.(2015)]{ale15} Alexander, K.D., Soderberg, A.M., \& Chomiuk, L.B., 2015, \apj, 806, 106
    
    \bibitem[Bietenholz et al.(2008)]{bie08} Bietenholz, M., 2008, in The role of VLBI in the Golden Age for Radio Astronomy, p.64 
    (Online at http://pos.sissa.it/cgi-bin/reader/conf.cgi? confid=72)
    
    \bibitem[Bietenholz et al.(2003)]{bie03} Bietenholz, M., Bartel, N., Rupen, M. P., 2003, \apj, 597, 374
    
    \bibitem[Bietenholz et al.(2011)]{bie11} Bietenholz, M., Bartel, N., Rupen, M. P. et al., 2011, in 
    VLBI and the New Generation of Radio Arrays, p.57 (Online at http://pos.sissa.it/cgi-bin/reader/conf.cgi? confid=125)
    
    \bibitem[Bietenholz et al.(2012)]{bie12} Bietenholz, M., Brunthaler, A., Soderberg, A.M., et al., 2012, \apj, 751, 125
    
    \bibitem[Bj\"{o}rnsson(2013)]{bjo13} Bj\"{o}rnsson, C.-I., 2013, \apj, 769, 65
    
    \bibitem[Bj\"{o}rnsson(2015)]{bjo15} Bj\"{o}rnsson, C.-I., 2015, \apj, 813, 43
    
    \bibitem[Caprioli \& Spitkovsky(2014a)]{c/s14a} Caprioli, D., \& Spitkovsky, A., 2014a, \apj, 783, 91
    
    \bibitem[Caprioli \& Spitkovsky(2014b)]{c/s14b} Caprioli, D., \& Spitkovsky, A., 2014b, \apj, 794, 46
    
    \bibitem[Caprioli \& Spitkovsky(2014c)]{c/s14c} Caprioli, D., \& Spitkovsky, A., 2014c, \apj, 794, 47
    
    \bibitem[Caprioli et al.(2015)]{cap15} Caprioli, D., Pop, A.-R., \& Spitkovsky, A., 2015, \apj, 798, L28
    
    \bibitem[Chevalier(1982a)]{che82a} Chevalier, R.A., 1982a, \apj, 258, 790
    
    \bibitem[Chevalier(1982b)]{che82b} Chevalier, R.A., 1982b, \apj, 259, 302
    
    \bibitem[Chevalier et al.(1992)]{che92} Chevalier, R.A., Blondin, J.M., \& Emmering, R.T., 1992, \apj, 392, 118
    
    \bibitem[Corsi et al.(2014)]{cor14} Corsi, A., Ofek, E.O., Gal-Yam, A., et al., 2014, \apj, 782, 42
    
    \bibitem[de Witt et al.(2016)] {dew16} de Witt, A., Bietenholz, M.F., Kamble, A., et al., 2016, \mnras, 455, 511
    
    \bibitem[Dickel(1991)]{dic91} Dickel, J.R., van Breugel, W.J.M., \& Strom, R.G., 1991, \aj, 101, 2151
    
    \bibitem[Fransson \& Bj\"{o}rnsson(1998)]{f/b98} Fransson, C., \& Bj\"{o}rnsson, C.-I., 1998, \apj, 509, 861
    
    \bibitem[Gal-Yam(2012)]{gal12} Gal-Yam, A., 2012, Science, 337, 927
    
    \bibitem[Guo et al.(2012)]{guo12} Guo, F., Li, S., Li, H., Giacalone, J., Jokipii, J.R., \& Li, D., 2012, \apj, 747, 98
    
    \bibitem[Gotthelf et al.(2001)]{got01} Gotthelf, E.V., Koralesky, B., Rudnik, L., Jones, T.W., Hwang, U., \& Petre, R., 2001, \apj, 552, L39
    
    \bibitem[Jun \& Norman(1996a)]{j/n96a} Jun, B.-I, \& Norman, M.L., 1996, \apj, 465, 800
        
    \bibitem[Jun \& Norman(1996b)]{j/n96b} Jun, B.-I, \& Norman, M.L., 1996, \apj, 472, 245
    
    \bibitem[Kasen \& Bildsten(2010)]{k/b10} Kasen, D., \& Bildsten, L., 2010, \apj, 717, 245
    
    \bibitem[Katsuda(2017)]{kat17} Katsuda, S., arXiv:1702.02054 [astro-ph]
    
    \bibitem[Krauss et al.(2012)]{kra12} Krauss, M.I., Soderberg, A.M., Chomiuk, l., et al., 2012, \apj, 750, L40
    
    \bibitem[Marcaide et al.(2009)]{mar09} Marcaide, J.M., Mart\'{i}-Vidal, I., Alberdi, A., et al., 2009, \aap, 505, 927
    
    \bibitem[Matheson et al.(2000)]{mat00} Matheson, T., Filippenko, A.V., Barth, A.J., et al., 2000, \apj, 120, 1487
    
    \bibitem[Park et al.(2015)]{par15} Park, J., Caprioli, D., \& Spitkovsky, A., 2015, \prl, 114, 085003
    
    \bibitem[P\'{e}rez-Torres et al.(2001)]{per01} P\'{e}rez-Torres, M.A., Alberdi, A., \& Marcaide, J.M., 2001,
    \aap, 374, 997
    
    \bibitem[Quimby et al.(2011)]{qui11} Quimby, R.M., Kulkarni, S., Kasliwal, M.M., et al., 2011, \nat, 474, 487
    
    \bibitem[Ressler et al.(2014)]{res14} Ressler, S.M., Katsuda, S., Reynolds, S.P., et al., 2014, \apj, 790, 85
    
    \bibitem[Reynoso et al.(2013)]{rey13} Reynoso, E.M., Hughes, J.P., \& Moffett, D.A., 2013, \apj, 145, 104
    
    \bibitem[Rybicki \& Lightman(1979)]{r/l79} Rybicki, G.B., \& Lightman, A.P., 1979, Radiative Processes in Astrophysics, Wiley, New York
    
    \bibitem[Ryder et al.(2004)]{ryd04} Ryder, S.D., Sadler, E.M., Subrahmanyan, R., Weiler, K.W., Panagia, N., \& Stockdale, C.,
    2004, \mnras, 349, 1093
   
    \bibitem[Schekochihin et al.(2004)]{sch04} Schekochihin, A.A., Cowley, S.C., Taylor, S.F., Maron, J.L., \& McWilliams, J.C., 2004,
    \apj, 612, 276
    
    \bibitem[Soderberg et al.(2010)]{sod10} Soderberg, A.M., Chakraborti, S., Pignata, G., et al., 2010, \nat, 463, 513
    
    \bibitem[Soderberg et al.(2006)]{sod06} Soderberg, A.M., Chevalier, R.A., Kulkarni, S.R., \& Frail, D.A., 2006, \apj, 651, 1005
    
    \bibitem[Soderberg et al.(2005)]{sod05} Soderberg, A.M., Kulkarni, S.R., Berger, E., et al., 2005, \apj, 621, 908
    
    \bibitem[Soderberg et al.(2012)]{sod12} Soderberg, A.M., Margutti, B., Zauderer, B.A., et al., 2012, \apj, 752, 78
    
    \bibitem[Tran et al.(2015)]{tra15} Tran, A., Williams, B.J., Petre, R., Ressler, S.M., \& Reynolds, S.P., 2015, \apj, 812, 101
    
    \bibitem[Weiler et al.(2011)]{wei11} Weiler, K.W., Panagia, N., Stockdale, C., Rupen, C., Sramek, R.A.,\& Williams, C.L.,   
    2011, \apj, 740, 79
    
    \bibitem[Weiler et al.(2007)]{wei07} Weiler, K.W., Williams, C.L., Panagia, N. et al., 2007, \apj, 671, 1959
       
\end{thebibliography}
\end{document}